\documentstyle[aps,prl,multicol]{revtex}
\include{psfig}
\begin{document}

\draft \title{
The phase space structure of supercooled liquids:
Evidence of a dynamical critical temperature.}

\author{Francesco Sciortino$^{1}$,
Srikanth Sastry$^{2}$ and Piero Tartaglia$^{1}$}

\address{$^{1}$ Dipartimento di Fisica and Istituto Nazionale
per la Fisica della Materia, Universit\'a di Roma {\it La Sapienza},
P.le Aldo Moro 2, I-00185, Roma, Italy. }
\address{$^{(2)}$ Department
of Chemistry Arizona State University Tempe, AZ 85287-1604 and Center
for Theoretical and Computational Materials Science National Institute
of Standards and Technology Materials Building 223/Room A313
Gaithersburg, MD 20899}

\date{\today}
\maketitle
\begin{abstract}
\noindent

We calculate local potential energy minima (inherent structures) for a
simple model of orthoterphenyl (OTP), from computer simulations over a
wide temperature range. We show that the results are very sensitive to
the system size. We locate, from the temperature and size dependence
of inherent structure energies, a well defined cross-over temperature
that corresponds to the dynamical critical temperature predicted by
mode coupling theory and in mean field spin glass models. Comparing
the known phase space structure of mean field $p$-spin models and the
present results for the OTP model, we find evidence supporting the
proposed similarity between glass-forming liquids and a class of spin
glass models.

\end{abstract}


\begin{multicols}{2}

The idea that essential features of glass formation from supercooled
liquids \cite{yip,angell} may be contained in the behavior of spin
glass models has been actively pursued in recent years, with a focus
on mean field $p$-spin models\cite{gros,kirk,parisi,cs,cavagna}. In
particular, the dynamical ideal mode coupling theory (MCT) for
supercooled liquids\cite{review-glass} and the disordered mean field
$p$-spin models, with $p>2$ display many common features observed in
structural glasses. Both approaches predict, on cooling the system,
the existence of a dynamical critical temperature ($T_{MCT}$ in MCT
language and $T_{D}$ in $p$-spin language) at which, from a dynamical
point of view, an ergodic to non-ergodic transition takes place. Below
this temperature the system is dynamically frozen in a metastable
state and confined in a finite region of phase space (basin, in the
following). Relaxation to the true equilibrium state below this
temperature takes place via activated processes, which are not taken
into account dynamically in MCT and are absent in the mean field
$p$-spin case, where all basins are separated by infinite barriers.
Both theories make predictions for the time dependence of correlation
functions (correlators). The evolution of correlation functions in the
$p$-spin case coincides \cite{kirk,cs} with that obtained for the
schematic MCT model\cite{schem}. In real structural glasses no sharp
ergodic to non-ergodic transition is observed, due to the finite
height of the barriers separating different basins, so that this
dynamical critical temperature plays the role of crossover temperature
between regions of different dynamical character. Specifically, it has
been argued that below this crossover temperature, activated crossing
of potential energy barriers, first suggested by Goldstein\cite{gold},
becomes the dominant mechanism of diffusion\cite{CAAtx}. MCT has been
the subject of several tests above $T_{MCT}$ and it has been shown to
describe the space and time dependence of correlators satisfactorily
\cite{kob,sgtc}.

On the other hand, mean field $p$-spin models have the merit of
offering a clear picture of the structure of phase space, in
particular the aspects associated with the transition at $T_{D}$. The
analytic solutions of the $TAP$ equations\cite{kirk,kurchan} show that
the region of phase space visited below $T_{D}$
is composed by an exponentially large
number of different basins, separated by infinitely high barriers.
Different basins are unambiguously specified by the value of the energy
at the minimum, $E_{min}$, which assume values only between a minimum
value $ E_o$ and a maximum value $ E_{th}$.  Stationary solutions with
$E_{min}$ higher than $ E_{th}$ are unstable even at zero temperature.
Below $T_D$, in thermodynamic equilibrium, basins with a well defined
value of $E_{min}(T)$ are populated at each temperature.

The $TAP$ solutions at $T = 0~K$ coincide with the local phase space
minima of the potential energy and are thus analog of the so-called
inherent structures proposed by Stillinger\cite{stillinger} to
describe the phase space structure of liquids. In Stillinger's
approach, phase space is partitioned in basins such that steepest
descent trajectories starting from any point in a basin converge to
the same energy minimum, which defines the inherent structure
configuration. Previous numerical studies of inherent structures in
supercooled liquids have been confined to rather small system sizes
and have rarely aimed at identifying the presence of an underlying
dynamical critical temperature. A recent study of a Lennard-Jones
binary mixture at constant density for a 256 atom sample\cite{sds} did
not reveal any particular signature of the presence of a dynamical
temperature in the $T$ dependence of the inherent structure energies.

This Letter explores the possibility that the phase space structure of
the liquid state is similar to that of $p$-spin models. The work is
based on a detailed study of the energy relaxation in phase space
toward the minimum, starting from equilibrated independent
configurations at a series of temperatures. The use of the conjugate
gradient technique\cite{numrec} guarantees an efficient monotonically
decreasing path toward the minimum and of course the absence of basin
change during minimization.  The basic idea behind this work is that
the equivalence between inherent structures of the liquid and TAP
solutions facilitates the comparison of the phase space structure
of  structural glasses and of mean field $p$-spin models.

We perform a series of long molecular dynamics (MD) simulations for
systems of different sizes ranging from 343 molecules (the smallest
size compatible with the intermolecular potential range) up to 9261
molecules\cite{simulations}. The selected intermolecular potential
models the OTP molecule as a rigid structure composed of three sites.
All site-site interactions are described by the same Lennard-Jones
potential, whose parameters have been optimized to reproduce the
properties of the liquid. For this potential $T_{MCT} = 280~K$, a
figure calculated from the $T$ dependence of the non-ergodicity
parameter\cite{lw}. Below $T_{MCT}$, the MD trajectories have been
calculated for times up to $20~ns$, to guarantee thermodynamic
equilibrium and satisfactory sampling of phase space. Both static and
dynamic quantities, evaluated along the MD trajectory, do not show any
finite size effect. Several different equilibrium configurations,
separated by more than the relaxation time have been used as starting
points for the evaluation of the corresponding inherent structures by
performing a standard conjugate gradient minimization. The energy
minimization has been performed up to a relative precision of
$10^{-12}$ to ensure a proper determination of the minimum.

Fig. \ref{fig:1} shows the potential energy during the minimization
procedure at different temperatures. While for $T \le 305~K $ there is
one single relaxation process to the minimum, with a characteristic
``time'' which has no system size dependence, at $T = 346~K $ the
relaxation to the minimum occurs in a two-step process.  The first
(size independent) relaxation process is followed by another slower
relaxation which drives the system to a deeper and size dependent
minimum. These data indicate that the system first relaxes quickly to
a quasi-stable size independent point characterized by an energy
$E_s(T)$ and then explores the possibility of finding a lower energy
minimum configuration with value $E_{min}(T)$. Strikingly enough, the
success of this search is strongly size dependent. Indeed, for the
smallest system $E_s(T)$ coincides with $E_{min}(T)$. It is
particularly important also to note that the change of behavior from
size dependent to size independent relaxation to $E_{min}$ occurs very
close to the previously estimated $T_{MCT}$.  In $p$-spin models, the
energy $E_{min}$ of the disordered solution (i.e. the stable solution
for $T \ge T_D$) is equal to that of the first (highest) stable $TAP$
solution.  By extrapolating $E_{min}(T)$ vs $1/N$ for $1/N$ going to
zero (see inset of Fig. \ref{fig:1}
we find that $\lim_{N\rightarrow\infty} E_{min}(T=346 K) $
is close to  $E_{min}(T=T_{MCT})$  which thus provides strong
evidence for the identification of $T_D$ with $T_{MCT}$.

In $p$-spin models, below $T_D$ the energies of the relevant $TAP$
minima lie below the threshold value $E_{th}$, and the curvature in
phase space of these $TAP$ solutions is positive in all directions.
Stationary solutions with energies greater than $ E_{th}$ display
unstable (or non-positive curvature) directions\cite{flat,cavagna}.
If this feature is preserved in the phase structure of supercooled
liquids, one would predict that, in the thermodynamic limit, for $T >
T_{MCT}$ there will be always a finite number of unstable directions
for higher energy stationary points and during a local energy
minimization, the system will always relax to $E_{th}$.  But if the
fraction of unstable directions $f_u$ is sufficiently small, as it is
close to $T_{MCT}$, in finite size systems
there would be realizations of higher energy
states which are stable.  For finite size systems, only when $6N
f_{u}(T)$ is greater than one, i.e. only when statistically there will
be at least one unstable solution, the system can abandon the saddle
state and relax toward a lower energy minimum. To test the validity of
this prediction, Fig. \ref{fig:2} shows the energy during the
minimization procedure at a $T$ close to $T_{MCT}$ for all studied
realizations of the large size system. For comparison, we show also
the relaxation curves averaged over all realizations for the 343
molecules cases.  Interestingly enough, a clear bimodal effect is
observed. A fraction of the realizations relaxes to $E_{min}$, while
the remaining fraction is trapped at the energy $E_s$. The smaller
system on average relaxes to $E_s$. As suggested by Fig. \ref{fig:1},
such system size dependence is seen only above the previously
estimated $T_{MCT}$ for this system.  The size effects displayed in
both figures, as well as the dichotomous behavior observed in
Fig.\ref{fig:2}, suggests that $T_{MCT}$ can be identified as the
temperature at which the fraction of unstable directions in phase
space $f_{u}(T=T_{MCT})$, evaluated at $E_s$, goes to zero from above.
In Ref. \cite{stprl}, by analysing the instantaneous normal mode
spectrum \cite{keyes} in a deep supercooled liquid above $T_{MCT}$, it
was shown that the fraction of truly unstable directions in phase
space tends to vanish at $T_{MCT}$, strongly suggesting that the
dynamics above and below $T_{MCT}$ is dominated by different physical
transport mechanisms: Above $T_{MCT}$ the dynamics is controlled by
the search of unstable directions while below it is
controlled by activated processes. We believe that our present
observations are relevant in establishing the nature of such a
dynamical crossover.

Fig. \ref{fig:3} shows the temperature dependence of $E_{s}(T)$ and
$E_{min}(T)$ for different system sizes.  $E_{s}(T)$ does not show any
significant dependence on the system size, suggesting that in
searching for the minimum the system quickly finds a local phase
space environment that does not depend on the system size. For $T$
higher than a size dependent cross-over temperature, a secondary
relaxation process appears -- suggesting that statistically at least
one direction in phase space at $E_s(T)$ is unstable, such that the
system relaxes to the deeper energy minimum $E_{min}(T)$. The cross
over temperature is size dependent and in the limit $1/N$ going to
zero extrapolates to $T \sim 285 K$ (see inset of Fig. \ref{fig:3}),
supporting the identification, in the thermodynamic limit, of this
cross over temperature with $T_{MCT}$. Above the cross-over
temperature there is a large $T$ range where $E_{min}(T)$ is
approximatively constant, suggesting that there is a well defined
minimum characteristic energy of the liquid state above $T_{MCT}$.
Below the crossover temperature the difference between $E_{s}(T)$ and
$E_{min}(T)$ disappears. No finite size effects are present. The phase
space visited in this $T$ range is partitioned in many different
basins. Interestingly enough, the variance in $E_{min}(T)$ is very
small below the crossover temperature, supporting the view that in the
thermodynamic limit, at each temperature, basins with a well defined
value of $E_{min}(T)$ are populated. It is also worth noting that
$E_s(T)$ is not only system size independent, but is an increasing
function of $T$ and coincides with $E_{min}$ up to the crossover
temperature for each system size.  The size effects shown in Fig.
\ref{fig:3} suggest the study of the phase space of the liquid not
only in terms of minima, but also in terms of stationary points of
higher order, as phase space features relevant to the system's
dynamics.  Finite dimensional cuts of the thermodynamic phase space
(finite $N$) transform saddle points into local minima, altering the
description of phase space as a more numerous collection of different
basins. In the thermodynamic limit, only the region of phase space
visited for $T < T_{MCT}$ can be described as a collection of
different basins, in agreement with the description of the free energy of
supercooled liquids proposed by Stillinger\cite{stillinger} and
observed in the $TAP$ solution of the $p$-spin models. Following the
analogy with disordered $p$-spin models, we note that $E_{min}(T)$
correspond to the equilibrium $TAP$ solutions.  Interestingly, the
finite size effect effects we report allow visualization of the
equivalent of the $TAP$ solutions for $E> E_{th}$, which have been
shown to be unstable stationary states\cite{flat}.

In summary, in this Letter we have presented results on the phase
space structure of supercooled liquids (structural glasses) and
discussed the analogies with disordered spin glasses. In particular
(i) we have identified $T_{MCT}$ in terms of a static property of the
system, showing that in the thermodynamic limit, $T_{MCT}$ corresponds
to the temperature at which the $T$ dependence of the inherent
structure energy has a clear break. This suggests a possible
experiment based on ultrafast quenches to determine the $T_{MCT}$ for
a liquid.  (ii) we have shown evidence that above $T_{MCT}$ the
inherent structure energy is $T$-independent\cite{highT}, suggesting
the possibility of a unique basin characterizing the liquid.  (iii)
Above $T_{MCT}$, size effects become important due to the
transformation of low dimensional saddle points to minima in finite
systems. The different structure of the phase space visited above and
below $T_{MCT}$ demonstrated by the present work offers precise hints
about the difference of dynamical behavior above and below
$T_{MCT}$. While below $T_{MCT}$ the slowing down of dynamics on
cooling can be connected to the time required to hop across energy
barriers between basins, above $T_{MCT}$ exploration of phase space
can take place {\it via} unstable directions connecting different
quasi-stable environments (defined by their value $E_{S}(T)$).  This
suggests further investigation of such quasi-stable solutions for
their relevance to the dynamics above $T_{MCT}$\cite{nature-note}.(iv)
Finally, based on the formal equivalence 
between $TAP$ solutions at $T = 0 K$ and
inherent structures, we have shown that the phase space structure of
the present model liquid is consistent with the phase structure of the
mean field $p$-spin models and have identified $T_{MCT}$ with $T_{D}$.

\noindent
\section{Acknowledgments}

We thank S. Ciuchi, B. Coluzzi, A. Crisanti, F. de Pasquale,
L. Fabbian, S. Franz, R.N. Mantegna,
M. Nicodemi, G. Parisi, G. Ruocco for stimulating
discussions. SS thanks Universit\'a di Roma {\it La Sapienza} for hospitality.

\end{multicols}

\begin{figure}
\caption{Relaxation of the potential energy per molecule during the
minimization procedure for three different system sizes, starting from
equilibrated configurations at three different temperatures $T=266$,
$T=305$ and $T=346$ K. Each curve is the average over several
independent minimizations.  Dotted line, $N=343$; Full line $N=1000$;
Long Dashed $N=9261$.
The inset shows the inherent structure
energy as a function of the inverse of the number of molecules $N$
for $T=305 K$ (dashed line) and for $T=346 K$ (full line).}
\label{fig:1}
\end{figure}

\begin{figure}
\caption{ Potential energy per molecule as a function of the number of
steps in the conjgate gradient procedure when $N=9261$. Different
curves refer to different equilibrium configurations at $T=318 K$. The
inset shows an enlarged view in the energy region around $E_s$ (full
lines) together with the averaged relaxation for $N=343$ (long
dashed).}
\label{fig:2}
\end{figure}

\begin{figure}
\caption{ Temperature dependence of $E_s$ (empty symbols) and
$E_{min}$ (filled symbols) as a function of temperature for three
different system sizes.  Square $N=9261$; Circle $N=1000$, Triangle
$N=343$. Lines ar drawn to guide the eye. It is important to note that
(i) below 300 K, $E_s$ coincides with $E_{min}$;
(ii) at $T=305 K$ finite size effects hinder the second relaxation process
also when $ N=9261$.}
\label{fig:3}
\end{figure}

\setcounter{figure}{0}

\eject

\begin{figure}
\centerline{\psfig{figure=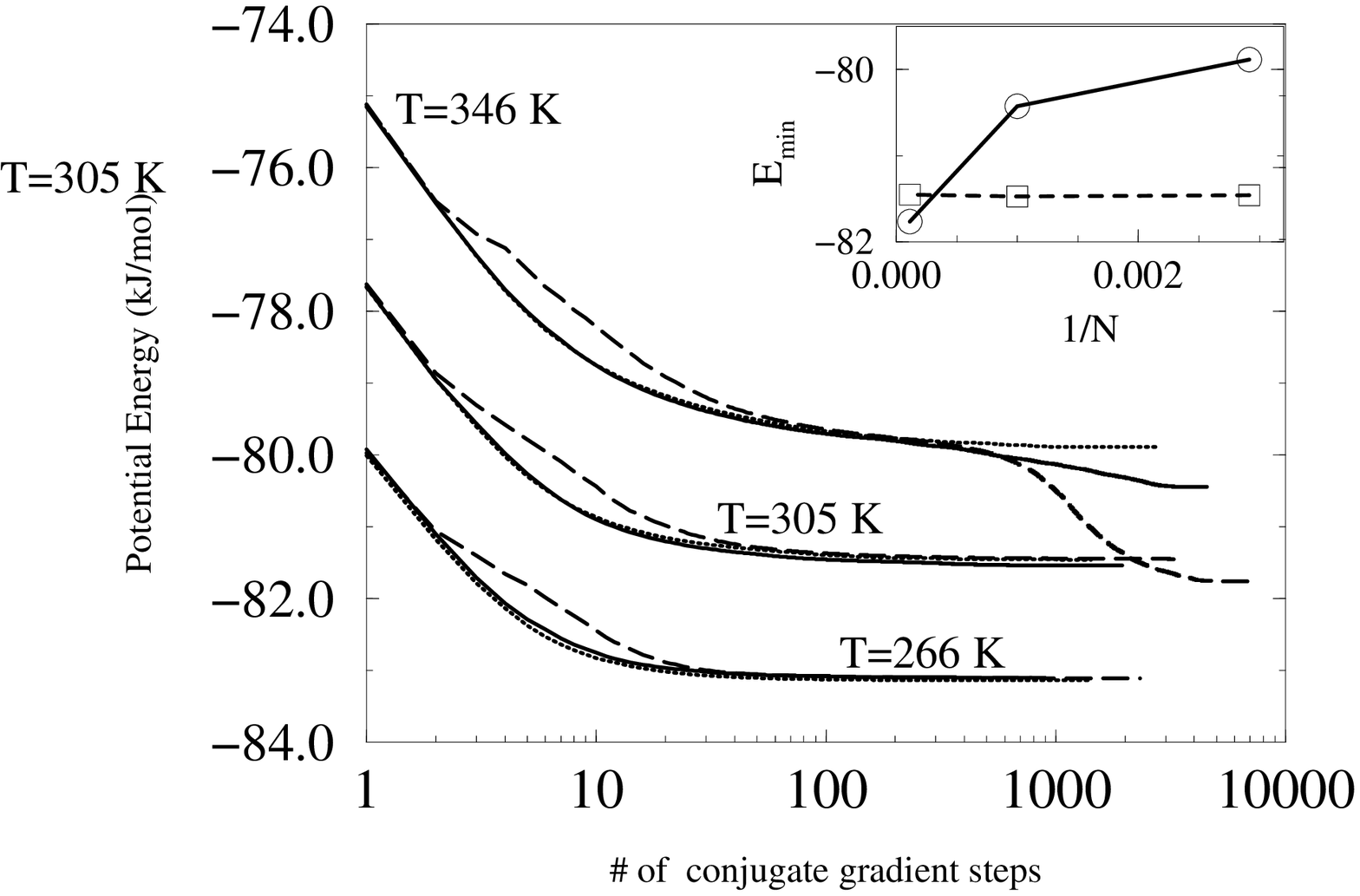,height=15cm,width=20cm,clip=,angle=0.}}
\caption{ F. Sciortino et al}
\end{figure}

\begin{figure}
\centerline{\psfig{figure=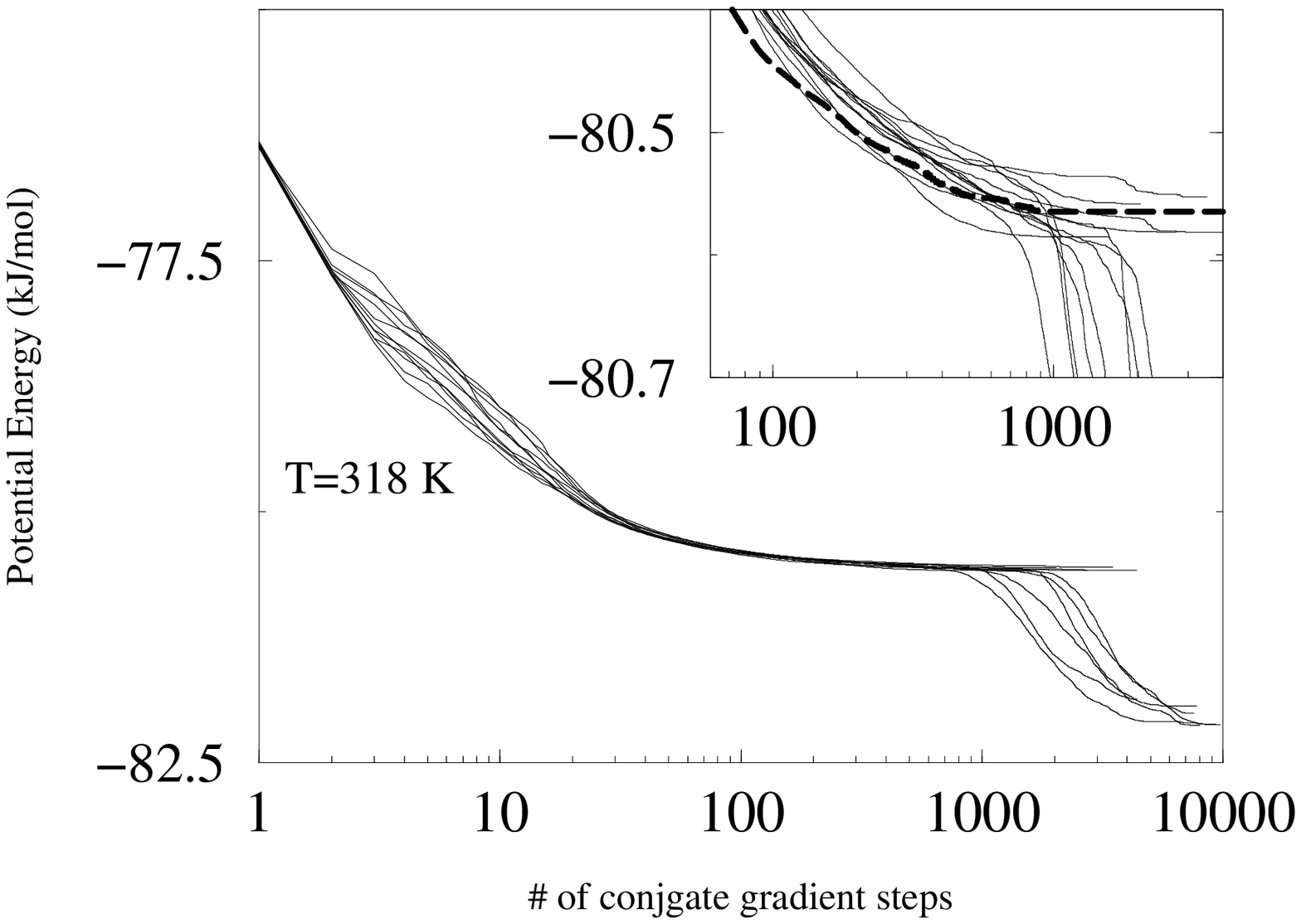,height=15cm,width=20cm,clip=,angle=0.}}
\caption{ F. Sciortino et al}
\end{figure}

\begin{figure}
\centerline{\psfig{figure=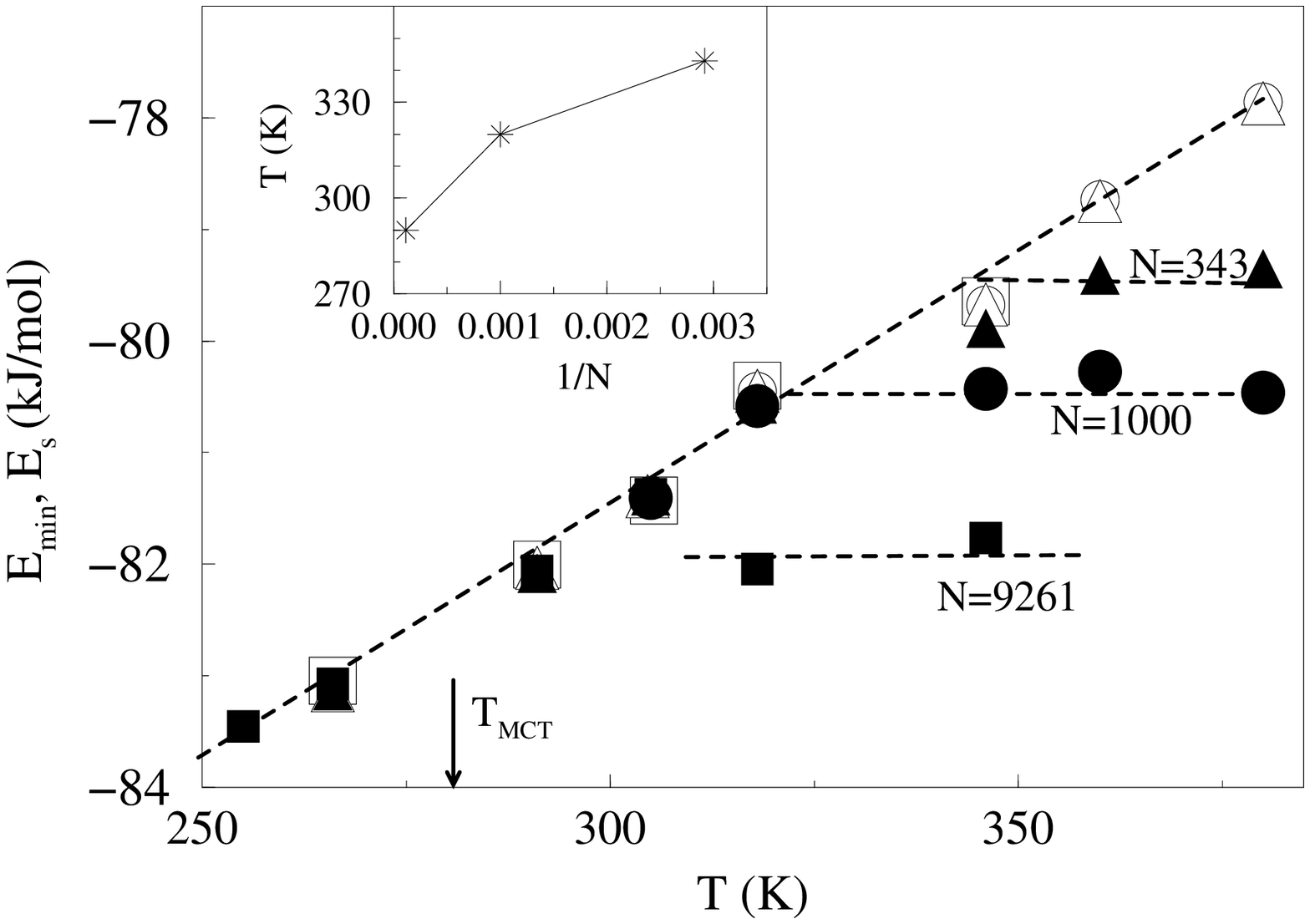,height=15cm,width=20cm,clip=,angle=0.}}
\caption{F. Sciortino et al}
\end{figure}

\begin{references}
\bibitem{simulations} We have simulated 12 different state points,
from $T = 255~K$ up to $T = 450~K$, following the constant ambient
pressure path previously studied in
Ref. \protect\cite{lw}. Simulations have been performed at constant
density. See table I in Ref. \protect\cite{lw} for the studied $T$ and
$\rho$.
\bibitem{highT} We plan to study in the near future the $T$ dependence
of the inherent structures for $T$ higher than the ones studied in
this Letter. This requires a careful study due to the vicinity of the
gas-liquid instability. We plan also to study a constant density path in the
phase diagram.
\bibitem{nature-note} Indeed, it appears plausible that
the onset of stretched
exponential and non-Arrhenius behavior above $T_{MCT}$ studied in
\cite{sds} is related to $E_s(T)$, which is under current
investigation.
\end{references}
\end{document}